\documentclass[aps,prd,groupedaddress,nofootinbib,superscriptaddress,preprint]{revtex4}
\usepackage{amsfonts}
\usepackage{amssymb}
\usepackage{amsmath,color}
\usepackage{epsfig}
\usepackage{graphicx,epsf,epsfig}
\usepackage{bbm}
\def\slc#1{\setbox0=\hbox{$#1$}           
    \dimen0=\wd0                                 
    \setbox1=\hbox{/} \dimen1=\wd1               
    \ifdim\dimen0>\dimen1                        
       \rlap{\hbox to \dimen0{\hfil/\hfil}}      
       #1                                        
    \else                                        
       \rlap{\hbox to \dimen1{\hfil$#1$\hfil}}   
       /                                         
    \fi}

\newcommand{\fig}{Fig.}
\newcommand{\eg}{\emph{e.g.}}
\newcommand{\ie}{\emph{i.e.}}
\DeclareMathOperator{\diag}{diag}

\begin{document}
\preprint{MPP-2010-6} \preprint{NORDITA-2010-2}
\title{Signatures from an Extra-Dimensional Seesaw Model}
\date{\today}
\author{Mattias Blennow}
\email{blennow@mppmu.mpg.de}

\affiliation{Max-Planck-Institut f{\"u}r Physik
(Werner-Heisenberg-Institut), F{\"o}hringer Ring 6, 80805
M{\"u}nchen, Germany}

\author{Henrik Melb{\'e}us}
\email{melbeus@kth.se}

\author{Tommy Ohlsson}
\email{tommy@theophys.kth.se}

\author{He Zhang}
\email{zhanghe@kth.se}

\affiliation{Department of Theoretical Physics, School of
Engineering Sciences, Royal Institute of Technology (KTH) --
AlbaNova University Center, Roslagstullsbacken 21, 106 91 Stockholm,
Sweden}
\begin{abstract}
We study the generation of small neutrino masses in an
extra-dimensional model, where singlet fermions are allowed to
propagate in the extra dimension, while the Standard Model
particles are confined to a brane. Motivated by the fact that
extra-dimensional models are non-renormalizable, we truncate the
Kaluza--Klein towers at a maximal Kaluza--Klein number.
This truncation, together with the structure of the bulk Majorana mass term, motivated by the
Sherk--Schwarz mechanism, implies that the Kaluza--Klein modes of the singlet fermions
pair to form Dirac fermions, except for a number of unpaired
Majorana fermions at the top of each tower. These heavy Majorana
fermions are the only sources of lepton number breaking in the
model, and similarly to the type-I seesaw mechanism, they naturally
generate small masses for the left-handed neutrinos. The lower Kaluza--Klein
modes mix with the light neutrinos, and the mixing effects are not
suppressed with respect to the light-neutrino masses. Compared to
conventional fermionic seesaw models, such mixing can be more significant. We study the
signals of this model at the Large Hadron Collider, and find
that the current low-energy bounds on the non-unitarity of the leptonic mixing matrix are strong
enough to exclude an observation.
\end{abstract}
\maketitle
\section{Introduction} \label{sec:intro}

Experimental studies of neutrino oscillations have provided us with
compelling evidence that neutrinos have masses and lepton flavors
mix. Among various theoretical models, the famous seesaw mechanism
\cite{Minkowski:1977sc,Yanagida:1979as,Mohapatra:1979ia,GellMann:1980vs} provides us
with a very natural description of why the masses of the three known
neutrinos are so small compared to the masses of the other Standard Model
(SM) fermions. In the simplest type-I seesaw model, heavy right-handed
neutrinos with a mass scale $M_{\rm R}$ are introduced in addition to the
SM particle content. In order to stabilize the masses of the light
neutrinos around the sub-eV scale, $M_{\rm R} \sim 10^{14}
~{\rm GeV}$ is naturally expected, if the Dirac mass
$m_{\rm D}$ between the left- and right-handed neutrinos is comparable with the mass of the top quark. The testability of
conventional seesaw models is therefore questionable. Furthermore,
the heavy right-handed neutrinos potentially contribute to the
hierarchy problem through loop corrections to the Higgs
potential, unless a supersymmetric framework is considered.

The Large Hadron Collider (LHC) will soon start to probe TeV scale
physics, and the question of whether we can find hints on the
neutrino mass generation mechanism at the LHC or not is relevant and
interesting. There are several indications that new
physics will show up at the TeV scale, in particular theories that
are able to stabilize the Higgs mass and to solve the
gauge hierarchy problem. The geometric mean of the Planck mass and
the $2.7$ K background temperature also suggests that 1 TeV is the
maximum mass that any cosmologically stable perturbatively coupled
elementary particle can have, otherwise the density of the Universe
exceeds its critical value \cite{Dimopoulos:1990gf}. Within the
seesaw framework, for the purpose of lowering the seesaw scale
without spoiling the naturalness criterion, some underlying symmetry
preserving the lepton number, $L$, is usually incorporated. For
example, in the type-I seesaw with more than one heavy right-handed
neutrino, contributions to the light-neutrino masses from different
right-handed neutrinos may cancel each other due to the symmetry, which results in massless left-handed neutrinos after
integrating out the heavy degrees of freedom from the theory
\cite{Kersten:2007vk}. Such a low-scale fermionic seesaw mechanism may not be
able to stabilize the masses of the light neutrinos, since loop corrections
may be unacceptably large. A
possible way to avoid this problem of the type-I seesaw model is
given by the inverse seesaw model, which
contains a Majorana insertion used to reduce the $B-L$ scale
\cite{Schechter:1980gr}. In the type-II seesaw model, extending the
SM with an $SU(2)$ triplet Higgs scalar
\cite{Schechter:1980gr,Lazarides:1980nt,Mohapatra:1980yp}, the
coupling between the triplet and the SM Higgs scalar breaks lepton
number explicitly and is expected to be very small. Thus,
the masses of the light neutrinos are suppressed through the approximate symmetry. In general, the canonical leptogenesis mechanism
\cite{Fukugita:1986hr}, which provides a very attractive description of the origin
of the observed baryon asymmetry of the Universe, does not work for
the low-scale seesaw mechanisms unless severe fine-tuning is invoked \cite{Pilaftsis:2005rv}.

In this paper, we employ the alternative framework of extra
spacetime dimensions, where the fundamental Grand Unified scale and
the Planck scale are lowered in a natural way
\cite{ArkaniHamed:1998rs}. We work exclusively within the context of flat extra dimensions. In our higher-dimensional
seesaw model, a truncating scale restoring the renormalizability of
the theory plays the role of breaking $B-L$, so that the masses of the light neutrinos are suppressed, while the lower Kaluza--Klein (KK)
states can be searched for at the LHC. Significant low-energy non-unitary leptonic mixing, due to integrating out the heavy
KK states, could give observable phenomena in future neutrino
oscillation experiments, such as a neutrino factory
\cite{Campanelli:2002cc,FernandezMartinez:2007ms,Altarelli:2008yr,Goswami:2008mi,Antusch:2009pm}.
In addition, resonant leptogenesis could possibly be achieved in this model. For earlier studies of the generation of small neutrino masses in the context of extra dimensions, see for example Refs.~\cite{Dienes:1998sb,Park:2009cm,Grossman:1999ra}. A study of unitarity violation in scenarios with bulk gauge singlet neutrinos was performed in Ref.~\cite{Bhattacharya:2009nu}. An alternative higher-dimensional seesaw model was investigated in Ref.~\cite{Frere:2003hn}.

The remaining parts of the paper are organized as follows: First, in
Sec.~\ref{sec:introduction}, we
present the general formalism of our model. Then, in Sec.~\ref{sec:NU}, we
show explicitly how sizable non-unitarity effects emerge in the
leptonic flavor mixing. Section \ref{sec:LHC} is devoted to
the collider signatures and the discovery potential of the heavy KK
modes at the LHC. We comment on the origin of
baryon number asymmetry in our model in Sec.~\ref{sec:leptogenesis}.
Finally, a summary and our conclusions are given in Sec.~\ref{sec:summary}.

\section{Higher-dimensional seesaw model} \label{sec:introduction}

We consider a brane world theory with a five-dimensional bulk, where
the SM particles are confined to the brane. We also introduce three
SM singlet fermions $\Psi_i$ ($i=1,2,3$)
\cite{Dienes:1998sb,ArkaniHamed:1998vp,Dvali:1999cn,Barbieri:2000mg,Lukas:2000rg}.
Being singlets, they are not restricted to the brane and can propagate in the
extra spacetime dimensions. The action responsible for the neutrino masses is given by
\begin{eqnarray}\label{eq:S5}
S & = & \int {\rm d}^4 x {\rm d}y \left[ {\rm i}\overline{\Psi}
\slc{D} \Psi - \frac{1}{2} \left(\overline{\Psi^c} M_{\rm R} \Psi +
{\rm h.c.} \right) \right] \nonumber \\
&  &
+\int_{y=0} {\rm d}^4 x \left( - \frac{1}{\sqrt{M_S}}
\overline{\nu_{\rm L}} \hat m^c \Psi - \frac{1}{\sqrt{M_S}}
\overline{\nu^c_{\rm L}} \hat m \Psi + {\rm h.c.}\right),
\end{eqnarray}
where $y$ is the coordinate along the extra compactified dimension and $M_S$ denotes the mass scale of the higher-dimensional theory. Note that, although $\Psi^c$ is defined in the same way as in four dimensions, it does not represent the charge conjugate of $\Psi$ in five dimensions \cite{Pilaftsis:1999jk}, and hence, the term $\overline{\Psi^c} M_{\rm R} \Psi$ is not a Majorana mass term\footnote{Majorana mass terms are not allowed in five-dimensional spacetime \cite{Weinberg:1984vb}.}. However, in the four-dimensional theory, it leads to effective Majorana mass terms for the KK modes of $\Psi$. Due to the freedom in the choice of basis for the singlet fermion fields,
one can always apply a unitary transformation in flavor space in
order to diagonalize $M_{\rm R}$. Without loss of generality, we
will therefore work in a basis in which $M_{\rm R} =
\diag(M_1,M_2,M_3)$ is real and diagonal. The Dirac masses $\hat m$ and $\hat m^c$ could be generated by couplings of the bulk neutrinos to a brane-localized Higgs boson receiving a vacuum expectation value.


We decompose the spinors of the bulk singlet fermions into two two-component objects: $\Psi=(\xi \,\, \eta^c)^T$, where $\eta^c = {\rm i} \sigma^2 \eta^*$. Since the extra dimension is compactified on the $S^1/\mathbb{Z}_2$ orbifold, the KK modes of $\xi$ and $\eta^c$ are four-dimensional Weyl spinors. We take $\xi$ to be even under the ${\mathbb Z}_2$ transformation $y\to-y$, while $\eta$ is taken to be odd. Thus, in Eq.~\eqref{eq:S5}, the $\hat m^c$ term corresponding to the coupling between $\nu_L$ and $\eta$ is not allowed. The KK expansions of $\xi$ and $\eta$ are given by
\begin{eqnarray}\label{eq:expand}
\xi(x,y) & = & \frac{1}{\sqrt{\pi R}}\xi^{(0)}(x) + \sqrt{\frac{2}{\pi R}} \sum^N_{n=1} \xi^{(n)}(x)
\cos\left(\frac{ny}{R}\right), \nonumber \\
\eta(x,y) & = & \sqrt{\frac{2}{\pi R}} \sum^N_{n=1} \eta^{(n)}(x)
\sin\left(\frac{ny}{R}\right).
\end{eqnarray}
In general, an extra-dimensional model must be viewed as an effective theory, since it is non-renormalizable. This means that the KK towers are expected not to be infinite, but truncated after a finite number of levels. The nature of this cutoff depends on the specific ultraviolet (UV) completion of the model, which is not known. Here, we impose a truncation of the KK towers at a maximum KK index $n=N$. A cutoff of this kind arises, for example, in deconstructed models of extra dimensions \cite{ArkaniHamed:2001ca}. In general, other kinds of truncation schemes are possible, but the one that we consider has the virtue of giving rise to a mechanism for generating small neutrino masses from the tops of the KK towers, as will be discussed below.

Inserting the above expansion into Eq.~\eqref{eq:S5} and
integrating over the compactified extra dimension, we arrive at the
following form for the four-dimensional action
\begin{eqnarray}\label{eq:S4}
S & = & \int {\rm d}^4 x \left\{{\xi}^{(0)\dagger} {\rm i}
\bar{\sigma}^\mu
\partial_\mu \xi^{(0)} +  \sum^N_{n=1} \left(
{\xi}^{(n)\dagger} {\rm i} \bar{\sigma}^\mu
\partial_\mu \xi^{(n)} +
{\eta}^{(n)\dagger} {\rm i} \bar{\sigma}^\mu
\partial_\mu \eta^{(n)} \right) \right. \nonumber \\
&& \phantom{\int d^4 x}-\frac{\rm i}{2}\left[ \left.  {\xi^{(0)}}^T
\sigma^2 M_{\rm R}  \xi^{(0)} + \sum^N_{n=1} \left(
\begin{matrix} {\xi^{(n)}}^T & {\eta^{(n)}}^T
\end{matrix} \right) \sigma^2 {\cal M}_n \left( \begin{matrix} {\xi^{(n)}}
\cr {\eta^{(n)}} \end{matrix} \right) + {\rm h.c.} \right] \right.
\nonumber
\\
&&
\phantom{\int d^4 x}-\left. {\rm i}  \left(\nu_{\rm L}^T \sigma^2 m_{\rm D} \xi^{(0)} + \sqrt{2}
\sum^N_{n=1} \nu_{\rm L}^T \sigma^2 m_{\rm D} \xi^{(n)} + {\rm h.c.} \right)
\right\},
\end{eqnarray}
where, written in block-form, the mass matrix ${\cal M}_n$ for the
KK modes at the $n$th level takes the form
\begin{eqnarray}\label{eq:MN}
{\cal M}_n = \left(\begin{matrix} M_{\rm R} & n/R \cr  n/R &
M_{\rm R}
\end{matrix}\right).
\end{eqnarray}
The Dirac mass term is then given by $m_{\rm D} = \hat
m /\sqrt{2\pi M_S R}$.

For the purpose of simplicity in the following discussion, we define
the linear combinations
\begin{eqnarray}\label{eq:chi}
X^{(n)} & \equiv & \frac{1}{\sqrt{2}} \left(\xi^{(n)} -
\eta^{(n)}\right), \nonumber \\
Y^{(n)} & \equiv & \frac{1}{\sqrt{2}} \left(\xi^{(n)} +
\eta^{(n)}\right),
\end{eqnarray}
for $n\geqslant 1$. The full mass matrix in the basis
$\left\{\nu_{\rm L},\xi^{(0)}, X^{(1)},Y^{(1)}, \cdots,
X^{(N)},Y^{(N)}\right\}$ then reads
\begin{eqnarray}\label{eq:M}
{\cal M} = \left(\begin{matrix} 0 & m_{\rm D} & m_{\rm D} & m_{\rm
D} & \cdots & m_{\rm D}  & m_{\rm D}  \cr m^T_{\rm D} & M_{\rm R} &
0 & 0 & \cdots & 0 & 0\cr m^T_{\rm D}  & 0 &\displaystyle   M_{\rm
R} - \frac{1}{R} & 0 & \cdots & 0 & 0 \cr m^T_{\rm D} & 0 & 0 &
\displaystyle M_{\rm R} + \frac{1}{R} & \cdots & 0 & 0 \cr \vdots &
\vdots & \vdots  & \vdots & \ddots & 0 & 0 \cr m^T_{\rm D} & 0 & 0 &
0 & 0 & \displaystyle   M_{\rm R} - \frac{N}{R} & 0 \cr m^T_{\rm D}
& 0 & 0 & 0 & 0 & 0 & \displaystyle   M_{\rm R} + \frac{N}{R}
\end{matrix}\right).
\end{eqnarray}
The scale of $M_{\rm R}$ is not governed by the electroweak symmetry
breaking, and hence, one can expect that $M_{\rm R} = {\cal O}(1)
\, {\rm TeV} \gg m_{\rm D}$ holds. Then, by approximately solving
the eigenvalue equation of the matrix in Eq.~\eqref{eq:M} with
respect to the small ratio $m_{\rm D}/M_{\rm R}$, the light-neutrino
mass matrix is found to be
\begin{eqnarray} \label{eq:mnuN}
m_\nu \simeq
 m_{\rm D}\left(\sum_{n=-N}^N \frac{1}{M_R + n/R}\right)m_{\rm D}^T
= m_{\rm D} \left(M^{-1}_{\rm R} + \sum^{N}_{n=1}
\frac{2M_{\rm R}}{M^2_{\rm R} - n^2/R^2} \right) m^T_{\rm D}.
\end{eqnarray}

In Refs.~\cite{Haba:2009sd,Bhattacharya:2009nu}, the limit $N \to
\infty$ is considered, and the light-neutrino mass matrix is then given by
\begin{eqnarray}\label{eq:mnu} m_\nu \simeq m_{\rm D}
\frac{\pi R }{\tan (\pi R M_{\rm R})} m^T_{\rm D}.
\end{eqnarray}
The masses of the light neutrinos are suppressed only if
$\tan (\pi R M_{\rm R})$ in the denominator of
Eq.~\eqref{eq:mnu} is very large. Therefore, a severe fine-tuning between $M_{\rm
R}$ and $R^{-1}$ has to be invoked, which appears quite unnatural.
However, bare Majorana masses of the form $M_{i} = k_i / (2 R)$,
where $k_i$ is an odd integer, emerge naturally from the
Sherk--Schwarz decomposition in string theory as a requirement of
topological constraints, and hence, such relations do not suffer any
fine-tuning problems (see detailed discussions in
Ref.~\cite{Dienes:1998sb}). With our chosen cutoff scheme, together with
the above condition on $M_i$, lepton number violation will be
induced only at the top of the KK tower, as we will see shortly.
There could, of course, be other lepton number violating processes
at some intermediate point, but we choose to treat the simple
scenario where the cutoff is the only source. One can easily prove
that, in the simplest case $k_i=1$, the light-neutrino mass
matrix is given by
\begin{eqnarray}\label{eq:m}
m_\nu \simeq m_{\rm D} \left(M_{\rm R} + \frac{N}{R}\right)^{-1}
m^T_{\rm D}.
\end{eqnarray}
Instead of a large mass scale $M_{\rm R}$ for the singlet fermions, the light-neutrino masses
are suppressed by the large cutoff scale $N/R$. We consider the interesting case
where the scale of the UV completion is much larger than the
scale of the extra dimension $1/R$ and the singlet fermion masses, \ie, we
assume $N \gg k_i$ to hold. In this limit, the neutrino mass matrix is simply given by $m_{\nu} \simeq
(R/N) m_{\rm D} m^T_{\rm D} $, \ie, the scale of the neutrino masses
is determined by a high-energy scale associated with the fundamental
theory underlying the effective extra-dimensional model. As for the
heavy KK modes, from Eq.~\eqref{eq:M}, the masses of the $n$th excited
KK modes are given by
\begin{eqnarray}
m_{X^{(n)}} & = & M_{\rm R} - \frac{n}{R}, \nonumber
\\ m_{Y^{(n)}} & = & M_{\rm R} + \frac{n}{R}.
\end{eqnarray}
As we will discuss later, this implies that $X^{(n)}$ and
$Y^{(n-1)}$ (as well as $X^{(1)}$ and $\xi^{(0)}$) form Dirac pairs.
Thus, lepton number can be assigned to these pairs and the lepton
number violating effects, such as neutrino masses, can only arise
from the unpaired $Y^{(N)}$ at the top of the KK tower.

\section{Non-unitary leptonic mixing}\label{sec:NU}

In order to compute the effective low-energy leptonic mixing, we
first consider the light-neutrino mass matrix. Generally,
$m_{\nu}$ is a complex symmetric matrix, and can be diagonalized by
means of a unitary matrix $U$ as
\begin{eqnarray}\label{eq:diag}
U^\dagger m_\nu U^* = D,
\end{eqnarray}
where $D={\rm diag}(m_1,m_2,m_3)$, with $m_i$ being the
masses of the light neutrinos. Note that, similarly to the ordinary fermionic
seesaw mechanism, the light neutrinos mix with the heavy KK modes. Thus, $U$
is not the exact leptonic mixing matrix entering into neutrino
oscillations, even if one works in a basis where the charged-lepton
mass matrix is diagonal. To see this point clearly, we can
fully diagonalize Eq.~\eqref{eq:M} and then write down the neutrino flavor
eigenstates in terms of the mass eigenstates
\begin{eqnarray}\label{eq:nu}
\nu_{\rm L} \simeq V \nu_{m{\rm L}} + K^{(0)} \xi^{(0)}+ \sum^N_{n=1}
\left[ K^{(-n)} X^{(n)} + K^{(n)} Y^{(n)}\right],
\end{eqnarray}
where $\nu_{m{\rm L}}$ denotes the mass eigenstates of the light
neutrinos, and $V$ is the upper-left $3\times3$ sub-matrix of the complete
mixing matrix containing the light neutrinos as well as the full KK tower for the singlet fermions.
Furthermore, we have introduced the quantities
\begin{equation}\label{eq:K}
K^{(n)}  =  { m_{\rm D} }{(M_{\rm R} + n/R)^{-1}},
\end{equation}
which represent the mixing between the light neutrinos and the KK modes. The
charged-current Lagrangian in mass basis can be rewritten as
\begin{eqnarray}\label{eq:Lcc}
{\cal L}_{\rm CC} = -\frac{g}{\sqrt{2}}   {\ell_{\rm L}^\dagger}
\bar{\sigma}^\mu \left[ V  \nu_{m{\rm L}} + K^{(0)} \xi^{(0)}+ \sum^N_{n=1}
\left( K^{(-n)} X^{(n)} + K^{(n)} Y^{(n)} \right) \right]W^-_\mu +
{\rm h.c.},
\end{eqnarray}
where $g$ is the $SU(2)$ coupling constant. Due to the existence of the KK modes, the light-neutrino mixing
matrix is no longer unitary. To a very good precision, we have
\begin{eqnarray}\label{eq:MNS}
V & \simeq &    \left(1-\frac{1}{2}\sum^N_{n=-N}
K^{(n)}{K^{(n)}}^\dagger
\right)U.
\end{eqnarray}
Assuming that $N\gg k_i$, Eq.~\eqref{eq:MNS} can
be approximated by
\begin{eqnarray}\label{eq:MNSapp}
V \simeq \left(1 - \frac{1}{2}\pi^2 R^2 m_{\rm D} m^\dagger_{\rm
D}\right) U.
\end{eqnarray}
Compared to the conventional parametrization of non-unitarity
effects $V=(1-\varepsilon)U$ \cite{Antusch:2006vwa}, where
$\varepsilon$ is a Hermitian matrix, we thus obtain
\begin{eqnarray}\label{eq:rho}
\varepsilon \simeq \frac{1}{2} \pi^2 R^2 m_{\rm D} m^\dagger_{\rm D}.
\end{eqnarray}
An interesting feature of Eq.~\eqref{eq:rho} arises immediately: the
non-unitarity effects are dominated only by the combination $m_{\rm
D}R$. As a rough estimate, if we keep $1/R$ at the TeV scale and
$m_{\rm D}\sim 100~{\rm GeV}$, $\varepsilon\sim 10^{-2}$ can be
naturally expected. Another typical feature is that, if $N \gg
k_i$ holds, then both the neutrino mixing and the non-unitarity
effects are determined by a single Dirac mass matrix $m_{\rm D}$. Therefore, in such a realistic low-scale extra-dimensional model, the
non-unitarity effects are strongly correlated with the neutrino
mixing matrix and the radius of the extra spacetime dimension. 

In our numerical computations, we adopt a convenient parametrization
\cite{Casas:2001sr}, and rewrite $m_{\rm D}$ as
\begin{eqnarray}\label{eq:MD}
m_{\rm D} = U \sqrt{D} O \sqrt{N/R},
\end{eqnarray}
with $O$ being an arbitrary complex orthogonal matrix. With this parametrization, Eq.~\eqref{eq:rho} takes the form
\begin{eqnarray}\label{eq:rhoA}
\varepsilon = \frac{1}{2} N \pi^2 R U \sqrt{D} O O^\dagger
\sqrt{D} U^\dagger \ .
\end{eqnarray}

The present bounds at 90~\% C.L. on the non-unitarity parameters are
given by \cite{Antusch:2006vwa,Antusch:2008tz}
\begin{eqnarray}\label{eq:rhoB}
|\varepsilon| < \left(\begin{matrix} 2.0\times10^{-3} & 6.0 \times
10^{-5} & 1.6 \times10^{-3} \cr \sim & 8.0\times10^{-4} & 1.1
\times10^{-3} \cr \sim & \sim & 2.7\times10^{-3}
\end{matrix}\right),
\end{eqnarray}
where the most severe constraint is that on the $e\mu$ element,
coming from the $\mu\to e\gamma$ decay. However, in the case that
$M_{\rm R}$ lies below the electroweak scale, but above a few GeV,
the $\mu\to e\gamma$ constraint is lost due to the restoration of
the Glashow--Iliopoulos--Maiani (GIM) mechanism \cite{Antusch:2008tz}, and a less stringent bound of
$|\varepsilon_{e\mu}| < 9.0 \times 10^{-4}$ should be used.

Apart from resulting in non-unitarity effects in neutrino
mixing, the heavy singlet fermions in the bulk will also contribute to
the lepton flavor violating (LFV) decays of charged leptons, \eg,
$\mu\to e\gamma$ and $\tau \to \mu \gamma$, through the loop
exchange of KK modes \cite{DeGouvea:2001mz}. Different from the
standard type-I seesaw mechanism, the corresponding branching ratios are not
dramatically suppressed by the light-neutrino masses, but only driven
down by the factor $K^{(n)}$ defined in Eq.~\eqref{eq:K}. Thus,
appreciable LFV rates could be obtained.

\section{Hadron collider signatures}\label{sec:LHC}

\begin{figure}[t]
\begin{center}\vspace{0cm}
  \includegraphics[width=16cm]{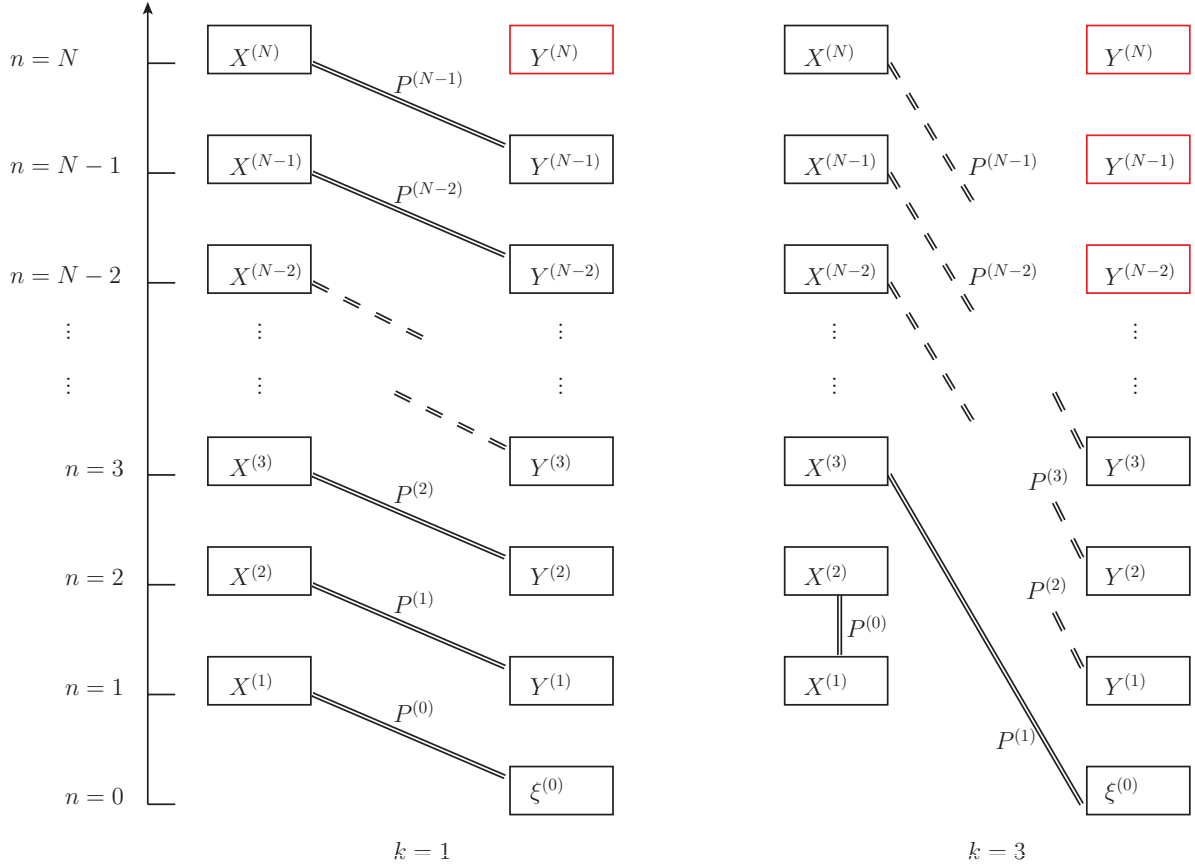}
\vspace{0.cm}
\caption{\label{fig:Dirac} Illustration of the
construction of Dirac particles from pairs of modes in the KK
tower. Two heavy KK Majorana modes with equal masses, but opposite CP
parities, can be grouped together, as shown with double lines, in order to form a
Dirac particle. In the case $k=1$ (left column), the heaviest mode
$Y^{(N)}$ is left, while for the case $k=3$ (right column), there
are three modes left: $Y^{(N-2)}$, $Y^{(N-1)}$, and $Y^{(N)}$.}
\end{center}
\end{figure}

As shown in Eq.~\eqref{eq:Lcc}, the heavy singlets $\xi^{(0)}$,
$X^{(n)}$, and $Y^{(n)}$ couple to the gauge sector of the SM, and
thus, if kinematically accessible, they could be produced at hadron
colliders. For a quantitative discussion, we now restrict ourselves
to the simplest case $k_i = 1$. Note that $\xi^{(0)}$ and $X^{(1)}$ are
two-component Majorana fields with equal masses but opposite CP
parities \cite{Bilenky:1987ty}. Thus, they are equivalent to a
single Dirac field $P^{(0)}$ with $P^{(0)}_L =\frac{1}{\sqrt{2}}
\left[ \xi^{(0)}+X^{(1)}\right]$, ${P^{(0)}_R}^c =\frac{1}{\sqrt{2}}
\left[ \xi^{(0)}-X^{(1)}\right]$, and mass $M_{P_0} = M_{\rm R}$.
Similarly, $X^{(2)}$ can be combined with $Y^{(1)}$, and hence,
forms a higher KK Dirac mode with $P^{(1)}_L = \frac{1}{\sqrt{2}}
\left[ Y^{(1)}+X^{(2)}\right]$ and mass $M_{P_1} = M_{\rm R} + 1/R$.
As a general result of the mass degeneracy, all the KK modes are
paired together except for the highest mode $Y^{(N)}$ with mass
$M_{\rm R}+N/R$. Actually, $Y^{(N)}$ is now the sole source of
lepton number violation, and thus, gives rise to the masses of the light
neutrinos, which can also be seen from Eq.~\eqref{eq:m}. The
structure of the singlet Dirac and Majorana fermions is schematically
depicted in \fig~\ref{fig:Dirac}.

The weak interaction Lagrangian for the heavy states can now be rewritten as
\begin{eqnarray}\label{eq:L}
{\cal L}_{\rm CC} & = & -\frac{g}{\sqrt{2}}   \ell_{\rm
L}^\dagger \bar{\sigma}^\mu \left[ V  \nu_{m{\rm L}} + \sqrt{2}\sum^{N-1}_{n=0}
K^{(n)} P^{(n)}_L +
K^{(N)} Y^{(N)} \right]W^-_\mu + {\rm h.c.}, \\
{\cal L}_{\rm NC} & = & \frac{g}{2 \cos \theta_{\rm W}} {\nu_{m\rm L}^\dagger}
\bar{\sigma}^\mu V^\dagger \left[ \sqrt{2} \sum^{N-1}_{n=0}
K^{(n)} P^{(n)}_L + K^{(N)} Y^{(N)} \right] Z_\mu+ {\rm h.c.}, \\
{\cal L}_{h} & = & \frac{-{\rm i} g}{\sqrt{2}M_{W}} {\nu_{m\rm L}^T} \sigma^2
V^T m_D \left[ \sqrt{2} \sum^{N-1}_{n=0} P^{(n)}_L
+  Y^{(N)} \right] h +
{\rm h.c.},
\end{eqnarray}
where $\theta_W$ denotes the weak mixing angle and
$M_W$ is the mass of the $W$ boson. In the case $M_{\rm R}>M_h$
(where $M_h$ denotes the Higgs mass), the heavy KK modes decay in
the channels $P \to \ell^- + W^+$, $P \to \nu + Z$, and $P \to h +
\nu$. The corresponding partial decay widths are given by
\cite{Buchmuller:1990vh}
\begin{eqnarray}
\Gamma\left(P^{(n)}_i \to \ell_\alpha W^+ \right) & = &
\frac{g^2}{32\pi} \left|K^{(n)}_{\alpha i}\right|^2
\frac{M_{P^{(n)}}^3}{M^2_W}
\left(1-\frac{M^2_W}{M_{P^{(n)}}^2}\right)\left(1+
\frac{M^2_W}{M_{P^{(n)}}^2} - 2\frac{M^4_W}{M_{P^{(n)}}^4} \right), \\
\nonumber \Gamma\left(P^{(n)}_i \to \nu_\alpha Z \right) & = &
\frac{g^2}{64\pi \cos^2 \theta_{\rm W}} \left|K^{(n)}_{\alpha i}\right|^2
\frac{M_{P^{(n)}}^3}{M^2_Z}
\left(1-\frac{M^2_Z}{M_{P^{(n)}}^2}\right)\left(1+
\frac{M^2_Z}{M_{P^{(n)}}^2} - 2\frac{M^4_Z}{M_{P^{(n)}}^4} \right), \\
\\
\Gamma\left(P^{(n)}_i \to \nu_\alpha h \right) & = &
\frac{g^2}{64\pi} \left|K^{(n)}_{\alpha i}\right|^2
\frac{M_{P^{(n)}}^3}{M^2_W} \left(1-\frac{M^2_h}{M_{P^{(n)}}^2}\right)^2,
\end{eqnarray}
where $M_Z$ and $M_{P^{(n)}}$ denote that masses of $Z$ and $P^{(n)}$, respectively.

\begin{figure}[t]
\begin{center}\vspace{0.cm}
\includegraphics[width=8cm]{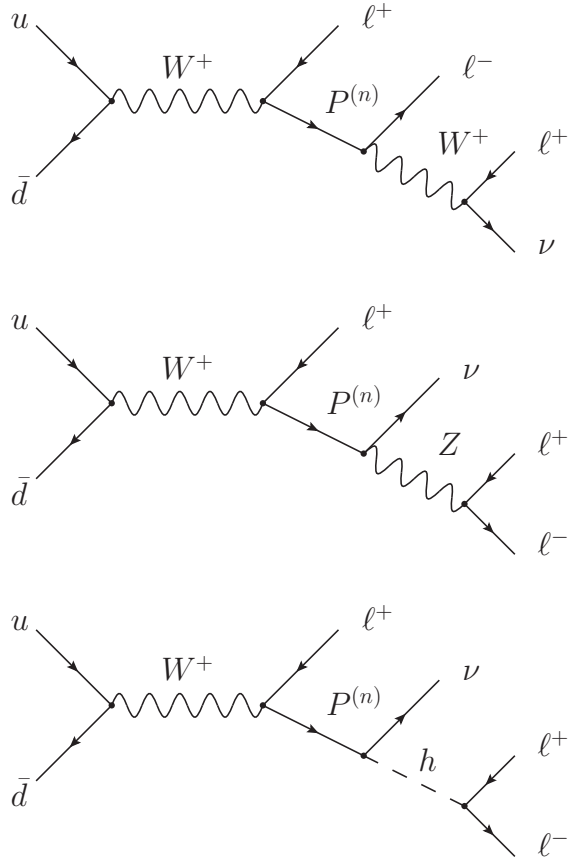}
\caption{\label{fig:LHC} Feynman diagrams for the potentially
interesting LHC signatures with three charged leptons and missing
energy in the model under consideration.} \vspace{-0.cm}
\end{center}
\end{figure}

Since the lower KK modes are Dirac particles, and lepton number
breaking occurs only at the top of the KK towers,
we focus our attention on lepton number conserving channels
mediated by the lightest KK modes. For example, an interesting
channel is the production of three charged leptons
and missing energy \cite{delAguila:2008hw}, \ie, $pp\to
\ell_{\alpha}^{\pm} \ell_{\beta}^{\pm}\ell_{\gamma}^{\mp}
\nu(\bar\nu)$, which is depicted in \fig~\ref{fig:LHC}.
Another possible process is the pair production of charged leptons
with different flavor and zero missing energy, \ie, $pp\to
\ell_{\alpha}^{\pm}\ell_{\beta}^{\mp}$. However, it is
difficult to make significant observations in this channel at the LHC,
due to the large SM background \cite{Aguila:2007em}.

An analysis of the collider signatures of an extra-dimensional model
similar to the one that we consider was performed in
Ref.~\cite{Haba:2009sd}. It was found that the most promising
channel for that model is three leptons and large missing energy. Since taus are difficult to detect,
due to their short lifetime, only electrons and muons in the final state were
considered. The signals were combined into two classes, the $2\mu$ signal, given by the sum of the $e \mu \mu$ and $\mu \mu \mu$ signals, where $e$ and $\mu$ denote both leptons and antileptons of the indicated flavors, and the $2e$ signal, given by the sum of the $e e \mu$ and $e e e$ signals. For the case of normal neutrino mass hierarchy ($m_3>m_2>m_1$), it was found that the $2\mu$ combination gives the most promising signal. In
order to reduce the SM background, which mainly comes from decays of
$Z$ bosons, the following kinematic cuts, taken from
Ref.~\cite{delAguila:2008cj}, were adopted: i) the two like-sign
leptons must each have a transverse momentum larger than 30 GeV and
ii) the invariant masses from the two opposite-sign lepton pairs
must each be separated from the mass of the $Z$ boson by at least 10
GeV. Only the effects of the lowest KK level were considered, as it
was concluded that the contributions from higher modes would be more
than one order of magnitude smaller.

\begin{figure}[t]
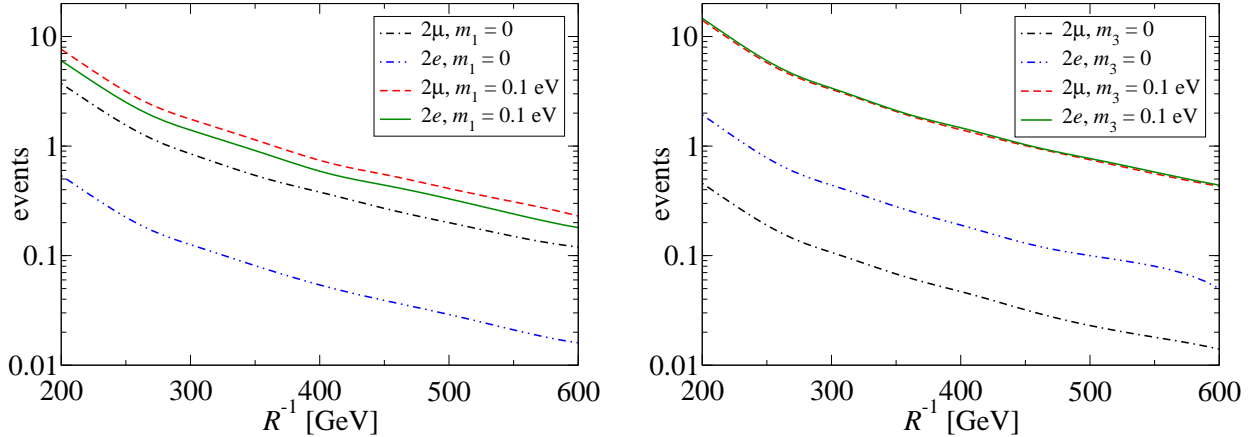

\begin{center}\vspace{0.cm}
\includegraphics[width=.48\textwidth]{normal.eps}\hspace{5mm}
\includegraphics[width=.48\textwidth]{inverted.eps}
\caption{\label{fig:LHCsignals} The expected number of events for the $2\mu$ and $2e$ signals at the LHC as functions of the inverse radius $R^{-1}$, for an integrated luminosity of $30~{\rm fb}^{-1}$. Note that the masses of the lightest singlet fermions are equal to $R^{-1}/2$. For $R^{-1} < 2 M_W$, on-shell production of the internal gauge bosons is not possible, and the signal is suppressed. The values of the neutrino oscillation parameters are given in the main text. Left panel: normal neutrino mass hierarchy. Right panel: inverted neutrino mass hierarchy.}
\end{center}
\end{figure}

We have calculated the $2\mu$ as well as the $2e$ signals for our model. The results, using an integrated luminosity of $30 \, {\rm fb}^{-1}$, are shown in Fig.~\ref{fig:LHCsignals}. We have considered the normal neutrino mass hierarchy ($m_3>m_2>m_1$) as well as the inverted hierarchy ($m_2>m_1>m_3$), and for each case, we have chosen the mass of the lightest neutrino to be equal to zero or $0.1$ eV, corresponding to the hierarchical or nearly degenerate neutrino mass spectrum, respectively. For the neutrino oscillation parameters, we have used the best-fit values from Ref.~\cite{Schwetz:2008er}, \ie, $\sin^2 \theta_{12} = 0.318$, $\sin^2 \theta_{13} = 0.013$, $\sin^2 \theta_{23} = 0.50$, $\Delta m_{21}^2 = 7.59 \times 10^{-5} \, {\rm eV}^2$, and $|\Delta m_{32}^2| = 2.40 \times 10^{-3} \, {\rm eV}^2$. We have put the Dirac CP-violating phase to zero. For each case, we have set the value of the cutoff scale in order to maximize the signal, while respecting the non-unitarity bounds given in Eq.~\eqref{eq:rhoB}. Like Ref.~\cite{Haba:2009sd}, we have only taken the lightest KK modes of the singlet fermions into account. The signals are dominated by the on-shell production of the internal gauge bosons and sterile fermions. Since $M_{P_1} = R^{-1}/2$, on-shell production of the gauge bosons is not possible if $R^{-1} < 2 M_W$, and in that case, the signals are suppressed by the off-shell propagators. Hence, we have chosen $R^{-1} = 200 \, {\rm GeV}$ as the lower bound in our figures.

In the case that the lightest neutrino is massless, the $2\mu$ signal is stronger than the $2e$ signal by approximately one order of magnitude for the normal hierarchy, while the opposite is true for the inverted hierarchy. In the case of a nearly degenerate mass spectrum, \ie, that the lightest neutrino has a non-zero mass equal to 0.1 eV, the two signals are almost identical, especially in the inverted hierarchy case. Since the expected background, after the kinematic cuts have been imposed, is of the order of 100 events \cite{delAguila:2008cj} and none of the signals is stronger than $\mathcal{O} (10)$ events, we conclude that, for our model, the non-unitarity bounds are strong enough to rule out the part of the parameter space that could possibly be probed by the LHC.

\section{Comments on leptogenesis}
\label{sec:leptogenesis}

Baryogenesis via leptogenesis is one of the main candidates for
being the theory appropriately describing the production of a baryon
asymmetry in the early Universe, which is measured to be $\eta_B =
(6.2\pm0.15)\times 10^{-10}$~\cite{Komatsu:2008hk}. In its most
basic form, leptogenesis occurs in a type-I seesaw scenario, where a
net lepton asymmetry is produced through the out-of-equilibrium
decay of the heavy neutrinos and then partially converted to a
baryon asymmetry through sphaleron processes. The Sakharov
conditions \cite{Sakharov:1967dj} are fulfilled by the decays
occurring out of equilibrium, the loop level CP-violation of the
decays through complex Yukawa couplings, and the baryon number
violation of the sphalerons, respectively.

Usually, the net lepton number is produced by the decays of the
lightest singlet fermions, since asymmetries produced by the
heavier neutrinos will be washed out. However, in our scenario, the
tower of Dirac fermions can be given definite lepton number
assignments and lepton number violation only occurs at the top of
the tower through the unpaired $Y$-states, which could take on the
role of the singlet fermions in the basic scenario. It is
important to note that for $k_i = 1$ ($i = 1,2,3$), there will be no
net lepton number violation, since all of the three unpaired states
will be degenerate in mass. However, if the $k_i$ are different,
\eg, $k_1 = 3$ and $k_2$, $k_3 = 1$, then $Y^{(N-2)}_1$ (see
\fig~\ref{fig:Dirac}) will be the unique lightest Majorana state and
a net lepton asymmetry could be produced. Since the mass splitting
of $1/R$ between the $Y$-states is expected to be very small
compared to the masses, the model would have to be treated within
the framework of resonant leptogenesis \cite{Pilaftsis:2003gt}.
Furthermore, to accurately examine the prospects for leptogenesis in
this model, one would have to properly take into account the effects
of the Dirac tower. Even if the Dirac fermions in the tower
preserve lepton number, they do not participate in the sphaleron
processes, since they are SM singlets, which could hide some part of
the produced lepton number from the sphalerons if all Dirac
fermions do not decay before sphaleron processes become inactive.
Thus, a detailed analysis, which is beyond the scope of this paper,
would be required to properly analyze the prospects for leptogenesis
in this model.

For earlier studies of leptogenesis in extra dimensions, see for example Refs.~\cite{Pilaftsis:1999jk,Abada:2006yd}.

\section{Summary and conclusions}\label{sec:summary}

In this work, we have studied a possible mechanism for generating
small neutrino masses in the context of extra dimensions. In the
model that we consider, the SM particles are confined to a
four-dimensional brane, while three SM singlet fermions are allowed
to propagate in an extra dimension, compactified on the $S^1 /
{\mathbb Z}_2$ orbifold. Since extra-dimensional models are
generally non-renormalizable, and can only be considered as
effective theories, the KK expansions of the higher-dimensional
fields are expected to be truncated at some cutoff scale. We have
imposed a cut on the KK number, truncating the towers at $n=N$.

In the case that the bulk Majorana mass term for the singlet fermions has the form $M_R = k/(2R)$, where $k$ is an odd integer,
the KK modes of the singlet fermions pair to form Dirac fermions. Such a form for a Majorana mass is motivated by, for example, the
Scherk--Schwarz mechanism. Due to the
truncation of the KK towers, a number of unpaired Majorana fermions
remain at the top of each KK tower, and these are the only sources
of lepton number violation in this model. If the cutoff scale is
large, small masses for the left-handed neutrinos are naturally
generated.

Due to mixing between the light neutrinos and the KK modes of the
singlet fermions, large non-unitarity effects can be induced.
Since the masses of the light neutrinos are generated by the top of each
tower, these non-unitarity effects are not suppressed by the light-neutrino masses. Current bounds on the non-unitarity parameters have
constrained the parameter space of the model.

Finally, we have considered the prospects of observing the effects
of the lowest KK modes of the singlet fermions at the LHC.
In particular, we have considered the three leptons and large missing
energy signal, which has previously been found to be promising for a
similar model. We have found that, in contrast to the previous
results in the literature, the potential of discovering such models at the LHC is actually pessimistic. In
particular, the parts of the parameter space that could be probed
at the LHC are ruled out by the bounds imposed by the stringent
constraints on the effective low-energy leptonic mixing. However, the non-unitarity
effects in neutrino oscillations could be observable at future
neutrino factory experiments. Therefore, future long baseline
neutrino oscillation experiments could play a very complementary
role in searching for hints of extra dimensions.

\begin{acknowledgments}

We would like to thank Steve Blanchet for useful
discussions. We acknowledge the hospitality and support from the
NORDITA scientific program ``Astroparticle Physics
--- A Pathfinder to New Physics'', March 30 -- April 30, 2009 during
which parts of this study was performed.

This work was supported by
the European Community through the European Commission Marie Curie
Actions Framework Programme 7 Intra-European Fellowship: Neutrino
Evolution [M.B.], the Royal Swedish Academy of Sciences (KVA)
[T.O.], the G{\"o}ran Gustafsson Foundation [H.Z.], and the Swedish
Research Council (Vetenskapsr{\aa}det), contract no.~621-2008-4210
[T.O.].

\end{acknowledgments}


\end{document}